# DSR with Non-Optimal Route Suppression for MANETs

Boon-Chong Seet
bcseet@ieee.org

Bu-Sung Lee
ebslee@ntu.edu.sg

Chiew-Tong Lau
asctlau@ntu.edu.sg

School of Computer Engineering, Nanyang Technological University, Singapore

*This paper revisits the issue of route discovery in dynamic source routing (DSR) for mobile ad hoc networks (MANETs), and puts forward a proposal of a lightweight non-optimal route suppression technique based on the observation of a rarely noted but commonly occurring phenomenon in route discovery. The technique exploits the observed phenomenon to extract query state information that permits intermediate nodes to identify and suppress the initiation of route replies with non-optimal routes, even if the route query is received for the first time. A detailed evaluation of DSR with non-optimal route suppression is found to yield significant improvements in both protocol efficiency and performance.*

## I. Introduction

A MANET [2] is a wireless network operated entirely by mobile nodes that cooperate to setup a communication service with no support from wired networks required. Nodes in the wireless range of each other may communicate directly, while those far apart must rely on other nodes to route their messages for them. Routing in a MANET has been a challenging issue since the network topology may constantly change due to mobility, and the efficiency of routing is a major concern as energy and bandwidth in a MANET are premium resources.

Many routing protocols for MANETs have been proposed, which may be categorized into two general types: i) proactive; ii) reactive [3]. Proactive protocols [4, 5] attempt to maintain up-to-date information about the path from each node to every other node in the network through regular exchange of topology updates. Reactive protocols [6, 7] in contrast, are based on the on-demand philosophy that protocols discover and maintain paths to only destinations to which data packets must be sent. Resources such as energy and bandwidth are thus not expended to discover and maintain unneeded routes.

DSR [7] is a well studied reactive protocol, known to be one of the most efficient in terms of resource consumption in literature [8, 9]. The efficiency of DSR can be largely attributed to its aggressive caching that allows the effective suppression of query packets generated during route discovery. However, its gain in efficiency has been limited by the increase in number of route replies [10], in particular cached replies from intermediate nodes, due to a greater number of routes learned through the caching of overheard routing information. Thus, it is of interest to ask if the process of generating route replies can be also optimized so that the existing efficiency can be enhanced. For example, if a node processing a query has advance knowledge that its route to be returned is not optimal, e.g. in terms of hop count, then it may choose to not reply, even if the query is received for the first time. In current DSR and reactive routing in general, it is almost mandatory for an intermediate node to reply, if it knows a route to the destination of a query, it receives for the first time.

This paper devises a lightweight non-optimal route suppression technique for the generation of route replies, based on the observation of a rarely noted but commonly occurring phenomenon in route discovery. The technique exploits the observed phenomenon to extract query state information that permits intermediate nodes to identify and suppress the initiation of route replies with non-optimal routes, even if the route query is received for the first time. A detailed evaluation of DSR with non-optimal route suppression is found to yield significant improvements in both protocol efficiency and performance.

The rest of the paper is organized as follows. Section II describes the salient features of the DSR protocol. Section III reports the observed phenomenon and presents the technique for non-optimal route suppression in DSR. In Section IV, the performance evaluation methodology is given. The performance results are then presented and discussed in Section V. Finally, we conclude the paper in Section VI.

---

[†] Earlier version of this work can be found in [1]

## II. DSR

DSR [7] is a reactive routing protocol based on the concept of source routing [11] – a method whereby each packet carries the complete route (a series of nodes) to traverse from the source to destination. It consists of two main phases: route discovery and route maintenance.

Route discovery is performed when a source has a packet to send, but does not know a route to its destination. The source broadcasts a query called a Route Request (RREQ) to each of its immediate neighbor, which on receiving, checks whether it is the destination, or has a route to the destination. If so, the node unicasts a response called a Route Reply (RREP) back to the source, informing it of the route to the destination. The RREP follows a path that is typically the reverse of that followed by the RREQ. Otherwise, the node appends its own address to the RREQ and rebroadcasts the packet to its neighbors, which in turn process in the same manner. Each node only processes a given RREQ once and discards duplicates of the same RREQ received from its neighbors. Nodes detect a duplicate RREQ by tracking and comparing the ID and source address of each received RREQ. A RREQ having reached its hop-limit or maximum number of traversable hops upon arrival at a node will be automatically dropped. Once a RREP is received, the source immediately sends out the packet to its destination using the route obtained.

Route maintenance is then carried out on the route in use to detect any link breaks, e.g. due to node mobility. Each node on the current route is responsible for sensing whether the link to its next-hop is broken. If so, the node unicasts an error message called Route Error (RERR) back to the source, informing it of the link in error. Upon receiving, the source stops sending any more packets using the faulty route, and may initiate a route discovery to acquire a new route to the destination if no alternate route is readily available. In DSR, route maintenance may also involve allowing an intermediate node to send a gratuitous RREP to the source, informing it of a shorter route it detects during a packet's journey across the network.

In addition to the above features, a number of optimizations have also been proposed by the DSR authors, which we refer to as DSR's *native* optimizations. These include the following for optimizing the basic route discovery, which are relevant to this paper:

*Caching overheard routing information*: By virtue of source route and promiscuous listening, a node may overhear ongoing data transmissions nearby and cache routes carried in packets that are addressed to its neighbors. This increases the amount of routing information a node can learn and save for its future use, thus avoiding the overhead of route discovery.

*Non-propagating route requests*: A two-phase route discovery is introduced. In the initial phase, the source conducts a so-called *Ring Zero* search, in which a non-propagating RREQ that does not travel beyond 1-hop from the source is broadcast to query only its neighbors for the destination, or a route to the destination. If no RREP is received after a short timeout, a propagating RREQ that spans the network is transmitted. The prior use of non-propagating RREQ potentially incurs a slight delay of 1-hop round-trip time in exchange for a large saving on overhead packets by not having to flood in each route discovery.

*Preventing route reply storms*: A situation may arise where a group of nodes receiving a RREQ reply simultaneously from their caches, thereby creating a 'storm' that causes local congestion and packet collisions. To relieve this problem, DSR requires each node to defer its reply for a period proportional to the length of route in its RREP. If during this delay, the node hears a data packet using a route shorter than the one it is deferring, then it may infer that the source already has a better path to the destination, and may cancel its RREP for this route discovery.

## III. DSR with Non-Optimal Route Suppression

In this section, we first report the observation of a phenomenon in DSR's route discovery, and then present a technique that puts the observed phenomenon to good use for non-optimal route suppression in DSR.

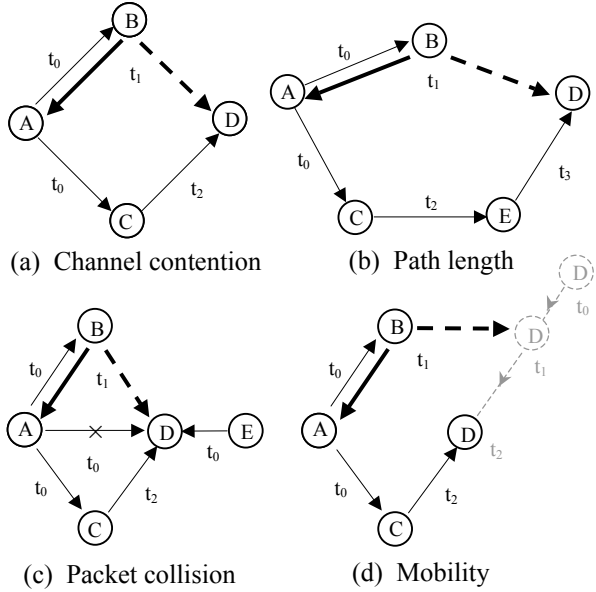

(a) Channel contention  (b) Path length

(c) Packet collision  (d) Mobility

Figure 1: Example scenarios that illustrate the observed phenomenon.

Referring to Figure 1(a), suppose a source *A* broadcast at time $t_0$ a RREQ, which is received by two intermediate nodes *B* and *C*. Assume *B* has a cached route to the RREQ destination, while *C* does not. Also assume *B* successfully contended for the channel and transmits ahead of *C* at time $t_1$ its Cached RREP, which is overheard by node *D*. After transmission from *B* is over, *C* rebroadcast the RREQ at time $t_2$, which is also received by *D*. Therefore, *D notices the RREP for a RREQ it receives only later*.

The above phenomenon also occurs in other scenarios. For example, Figure 1(b) illustrates a case where the RREQ travels over a longer path (*A-C-E*) than the RREP (*A-B*) to reach *D*. In Figure 1(c), *D* is in direct range of *A*, but did not receive its RREQ due to a packet collision caused by a simultaneous broadcast from *E* at time $t_0$ for a separate route discovery. Then as in Figure 1(a), *D* hears the RREP at time $t_1$ and receives RREQ at time $t_2$. Figure 1(d) further considers the effects of mobility, where *D* moves into range of *B*, and later into range of *C* at times $t_1$ and $t_2$ respectively, thereby observing the RREP *before* receiving the RREQ for the same route discovery.

It thus reverses a commonly assumed query (RREQ) precedes response (RREP) arriving order and this phenomenon can be put to good use to suppress the initiation of RREP, in particular Cached RREP with non-optimal routes from the intermediate nodes. For example, the overheard RREP provides information about i) the hop length of the returned source route, and ii) the ID of the RREQ, for which this RREP is initiated. Together with the source address of the RREQ, which is found from either the returned source route, or the destination address of the RREP, these three pieces of information can be used to decide if a node should reply upon receiving a RREQ, even for the first time.

Existing DSR and reactive routing in general dictates that a node should reply if it receives a RREQ for the first time, which it has a route to the destination. It is almost certain that node *D* in Figure 1 would send a Cached RREP, since it knows at least a route to the destination, which is that it overhears from the RREP. However, the returned route may not be useful if it is longer than the one previously returned, since route selection at the source is typically based on the shortest path.

We thus propose that if a node overhears a RREP for a RREQ it has not seen before (known by the RREQ ID and source address), the node shall record the three pieces of information from the RREP, namely i) the hop-length of returned source route, ii) RREQ ID, and iii) RREQ source address, as mentioned before. Subsequently, if the node receives this RREQ, it will compare the hop-length of its route (to be returned) with that seen previously. It will reply if it has a shorter route, and discard otherwise. Figure 2 shows the pseudo-code for this algorithm.

```
Let p be the RREP overheard from a neighboring transmission
Let q be the RREQ for which p is initiated
Let H_s be the hop-length of the returned source route s in p
Let H_r be the hop-length of the desired source route r in cache
Let ID be the identification value of a RREQ
Let S and D be the RREQ's source and destination respectively

// At intermediate nodes
When p is received:
    Check S and ID of q in p to determine if q has been seen
    If q has not been seen before
        Record H_s, S and ID of q
        Return

When q is received:
    Check route cache to determine if a route r to D exists
    If (r exists) AND (H_r < H_s)
        Send RREP with r to S
    Else   // if no route exists, or if r exists but H_r >= H_s
        Discard q   // do not forward or reply to q which
        Return      // has been replied with p previously
```

Figure 2: Route suppression algorithm

We recall that DSR has a scheme with some similarity for preventing reply storms (Section II). However, the scheme does not propose the use of other information received from the RREP as we mentioned above. In addition, the scheme listens for shorter routes only *after* RREQ is received, which introduces a delay that may increase the route acquisition latency.

## IV. Evaluation Methodology

### IV.A Simulation Model

The network simulator-2 (ns-2) [12] is used for evaluating the proposed technique in this paper. A total of 100 nodes are simulated for 500s over a network space of 1342m x 1342m. The traffic pattern is modeled as 40 CBR sources with data sent in 64-byte packets at 2 packets/s. We chose the above configuration as it allows a reasonably timed simulation, while stressing protocols with a sufficiently high load without causing congestion in the network.

Nodes move according to the random waypoint model [13], with pause times varying between 0 and 500 seconds. A pause time of 0s corresponds to continuous motion, while a pause time of 500s (length of simulation) corresponds to no motion. Each node may move at speed of up to 20m/s. We simulate each pause time with 5 movement scenarios, each generated using a different seed and present the mean of each performance metric over these 5 runs. The following describes our performance metrics of interest.

- *Route discovery overhead*. The sum of routing packets generated by each node due to route discovery. This includes both RREQ and RREP packets. Each hop-wise transmission is counted as one transmission.
- *Total routing overhead*. The sum of routing packets generated by each node due to both route discovery and route maintenance. The packets generated by basic route maintenance are mainly the RERR packets.
- *Route discovery latency*: The average time from originating a RREQ at the source to receiving the first RREP that answers the query. This metric measures the amount of time needed to acquire a route to the destination.
- *End-to-end delay*. The average time from originating a data packet at the source to delivering it to its intended destination. This includes all possible delays such as route discovery latency, queuing delay at network interface queue, propagation and retransmission delays in MAC and physical layers.
- *Packet delivery ratio*. The percentage of data packets delivered to the destinations with respect to number of data packets sent by the sources. This metric measures the extent of packet loss due to such reasons as routing failure and network congestion.

### IV.B Protocol Description

DSR+S is the combination of DSR [7] with the proposed route suppression. Its performance is compared with DSR, which is used with all its native optimizations to provide a challenging base for comparison in this paper.

DSR+S similarly performs a two-phase route discovery as DSR (Section II). The principal difference lies in the processing of received RREP and RREQ at intermediate nodes, which is modified to perform according to the pseudo-code given in Figure 2. Note that route suppression is not performed in every route discovery, but only during flooding in the second phase where the observed phenomenon occurs. Therefore its performance impact can be limited by the use of Ring Zero search during the initial phase. We shall return to this point when we discuss about the performance results in Section V. Table 1 summarizes the parameters used.

Table 1: Summary of parameters

| Parameter | Setting |
| --- | --- |
| Mobility model | Random waypoint |
| Traffic model | 40 CBR sources |
| Network space | 1340m x 1340m |
| Number of nodes | 100 nodes |
| Maximum node speed | 20 m/s |
| Packet sending rate | 2 packets/s |
| Data payload | 64 bytes |

## V. Performance Results

In the following section, we present and discuss the performance result for each metric separately. The results for all plots in the paper are shown with 95% confidence intervals.

### V.A Route Discovery Overhead

We first refer to Figure 3, which compares the number of RREP between DSR and DSR+S. The RREP are further segregated into Cached and Target RREP, the former being generated by non-destination (intermediate) nodes, while the latter by destination nodes. Due to source routing and aggressive caching, DSR has an inherent high hit rate for its route caches [10], which results in significantly more Cached RREP than Target RREP as shown in the figure.

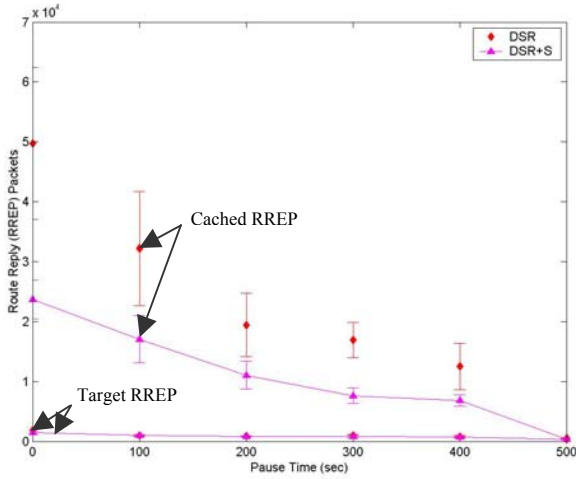

Figure 3: Comparison of number of RREP between DSR and DSR+S

The results above show that Cached RREP in DSR+S is significantly fewer, especially at higher node mobility (lower pause time). At pause time of 0s (highest mobility) where all nodes are in continuous motion, the number of Cached RREP is reduced by more than 50%. And expectedly, this margin of improvement decreases with mobility, since lower speed leads to fewer route discoveries to be performed. At pause time of 500s where all nodes are stationary, no significant difference in Cached RREP is observed. Also, since our scheme is aimed at Cached RREP, the number of Target RREP remains relatively unchanged. Figure 4 (a) and (b) shows the overall route discovery overhead, comprising both RREQ and RREP, of DSR and DSR+S, respectively. By taking RREQ into account, DSR+S achieves an overall reduction of 37.3% under the highest mobility (zero pause time).

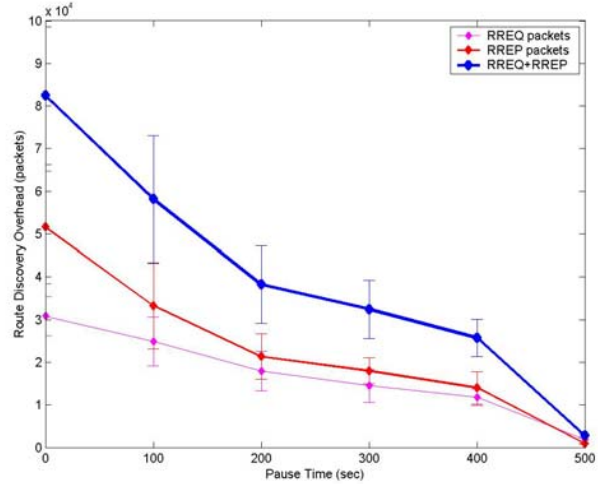

(a) DSR

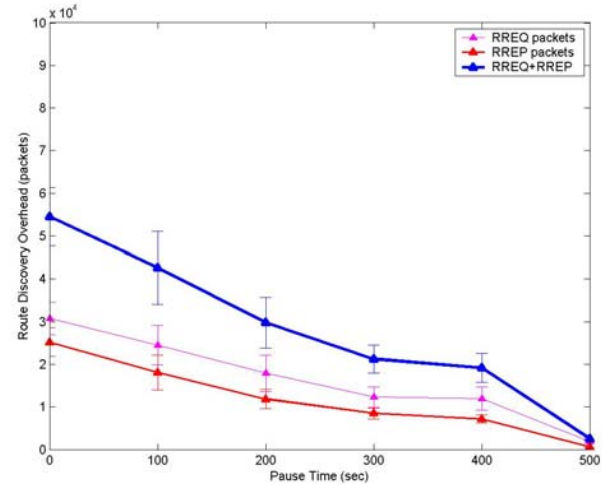

(b) DSR+S

Figure 4: Route discovery overhead

### V.B Total Routing Overhead

In this section, we determine the total routing overhead that also comprises route maintenance packets, which are packets sent to monitor or maintain routes in use. These are RERR packets, plus a small number of gratuitous RREP sent to optimize (shorten) the routes in use over time. Figure 5 shows the total routing overhead as a function of pause time.

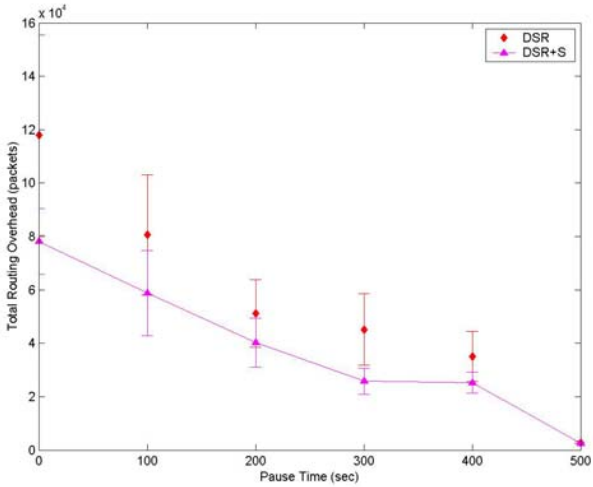

Figure 5: Total routing overhead vs. pause time

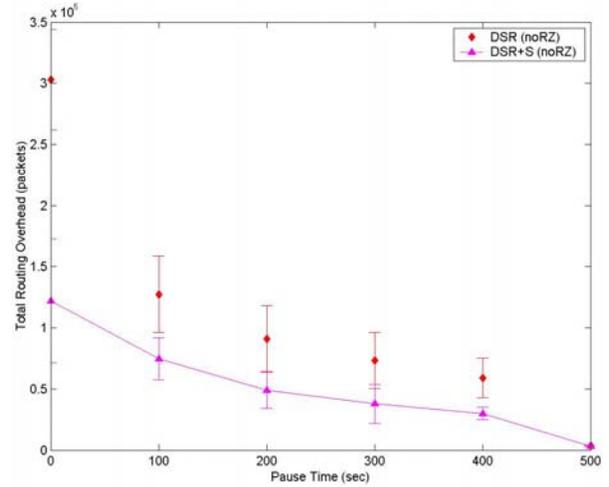

Figure 6: Total routing overhead with no Ring Zero incorporated

The result shows that the DSR's total routing overhead is reduced by as much as 37.5% under highest mobility (pause time = 0s). This margin of improvement is close to that of route discovery (Section V.A) due to a similar reduction in the route maintenance packets. Although our scheme is targeting at Cached RREP, the suppression of RREP with longer routes reduces the potential for route breaks, which in turn reduces number of RERR. Also, by having less routing performed using longer routes, fewer gratuitous RREP are sent to optimize the route length.

Figure 6 further shows the overhead results of both protocols with no Ring Zero (noRZ) incorporated, i.e. each route discovery is performed via flooding. Using this protocol configuration is to evaluate the maximum improvement achievable by route suppression, which is otherwise limited by the success of Ring Zero, as briefly mentioned in Section IV.B. The results show that without Ring Zero, total routing overhead of DSR and DSR+S increases by as much as 2.5 and 1.5 times, respectively, while the maximum (relative) percentage improvement increases to 60%.

Figure 7 summarizes the overhead results and shows the composition of routing packets for each protocol under highest mobility (pause time = 0s). Note the proportion of Cached RREP (CREP) in the figure for DSR and DSR (noRZ). CREP is a significant component that accounts for almost half the sum of routing packets generated, which highlights the need to control transmission of this type of routing packet.

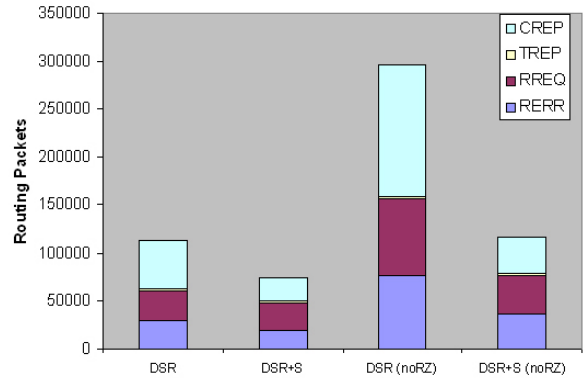

Figure 7: Composition of routing packets of each protocol

## V.C Route Discovery Latency

Figure 8 shows the route discovery latency as a function of pause time. In general, the latency increases with mobility. The variability in latency as observed from the confidence intervals, also increases with mobility, especially for DSR. This often suggests the presence of congestion, which adds random latencies to the forwarding delay of route discovery packets (RREQ and RREP). But as forwarding delay may also depend on the path taken by the packets to acquire a route, we also determine the average number of hops traversed by the first RREP received for a route discovery. However, as shown in Table 2, no major difference in the hop-count traversed is found. Thus, the observed latency difference between the protocols at higher mobility is mainly attributed to a higher forwarding delay due to routing loads.

Table 2: Average number of hops traversed by first received RREP

| Protocol | Hops traversed by first RREP |
|---|---|
| DSR | 2.80 |
| DSR+S | 2.90 |

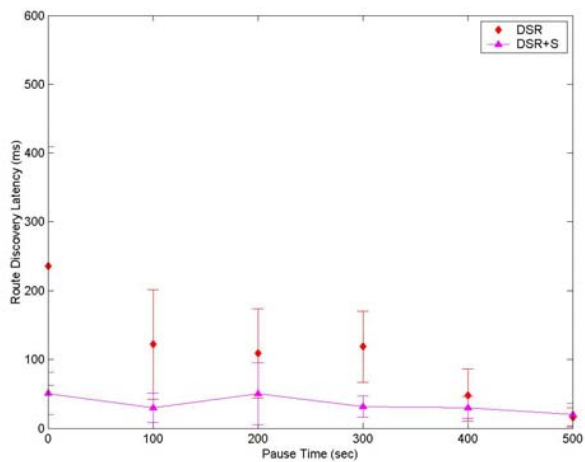

Figure 8: Route discovery latency vs. pause time

## V.D  End-to-End Delay

The end-to-end delay of a data packet includes the initial route discovery latency, and the subsequent forwarding delays it experiences along the route to its destination. From Figure 9, the result shows a similar trend as that for route discovery, since the congestion effect that impacts route discovery packets also impacts the data packets in a similar way. For DSR+S, a lower routing load means fewer routing packets compete with data packets for channel access, resulting in fewer collisions and backoffs that may prolong end-to-end delay. Given that the routing packets are often given a higher priority to transmit than data packets in interface queues, fewer routing packets also lead to shorter queuing time for data packets waiting to be transmitted.

Besides the congestion effect given above, the number of hops traversed by data packet to reach its destination may also impact end-to-end delay. Figure 10 shows the hop-wise optimality of the paths used in each protocol for packet delivery. It shows the path length difference between the actual number of hops taken and optimal number of hops required by data packets to reach their destinations. Here, a difference of zero means the data packet have taken an optimal (shortest) path, whereas a difference > 0, i.e. 1, 2 , 3, indicates the extra number of hops the packets have incurred. DSR+S is found to have slightly better path optimality than DSR, which may be attributed to its suppression of RREP with longer routes. But in general, the optimality of both protocols is fairly similar, with a large fraction of packets delivered using optimal (shortest), or near-optimal paths with one or few extra hops.

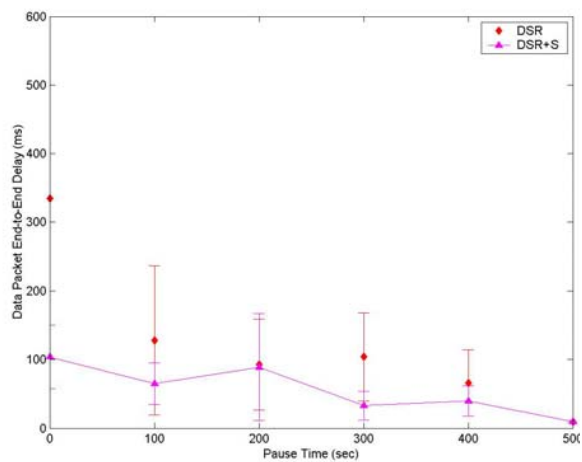

Figure 9: End-to-end delay vs. pause time

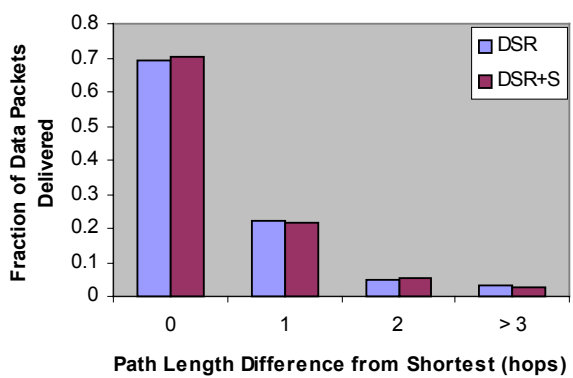

Figure 10: Path optimality of data packets

## V.E  Packet Delivery Ratio

The percent of data packets successfully delivered to their destination decreases with higher mobility (lower pause time), as shown in Figure 11. At the highest mobility (pause time = 0s), the packet delivery ratio of DSR decreases to below 90%, while that of DSR+S still remains above. This can be attributed in part to *reduced routing failure* and in part to *reduced congestion*, which can be observed from the reasons found for packet loss at highest mobility, as shown in Table 3. Fewer packets encountered routing failure (no route) in DSR+S, and as reflected from the fewer RERR in Figure 7, because longer routes are prevented from being returned and used for routing when route suppression is employed. The resulting lower routing load also contributes to reduced congestion, which in turn reduces the packet drops due to interface queue (IFQ) overflow.

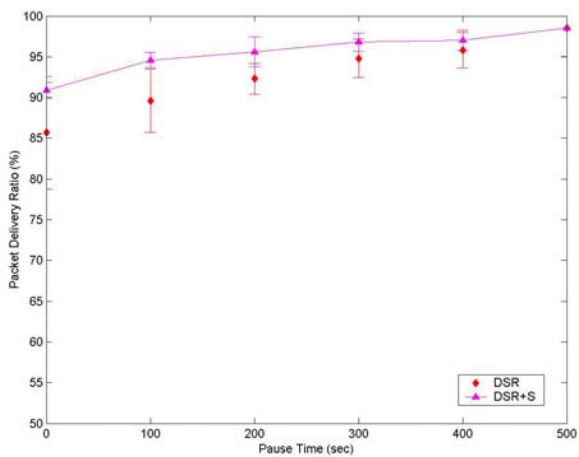

Figure 11: Packet delivery ratio vs. pause time

Table 3: Data packet dropped summary

| Data Packet Dropped Summary | DSR | DSR+S |
|---|---|---|
| No Route | 2401 | 2307 |
| TTL Expired | 0 | 0 |
| RTR Queue Full | 0 | 0 |
| Timeout | 137 | 141 |
| Routing Loop | 0 | 0 |
| IFQ Full | 757 | 36 |
| ARP Full | 138 | 153 |
| MAC Callback | 0 | 0 |
| Simulation End | 107 | 44 |
| Total Packets Dropped | 3540 | 2681 |

## V.F Effect of Network Size and Traffic Load

The previous sections investigated the effect of mobility on the protocols by varying the pause times for a given network size and traffic load. This section further investigates the impact of changing the network size and traffic load on the performance metrics. In order to isolate the effect of these changes, we chose to vary one parameter at a time, i.e. either network size or traffic load, but not both at the same time, while keeping other parameters unchanged. Network size is the number of nodes in the network, while traffic load is the number of data packets injected to the network, which can be varied by changing the number of data sources. Different numbers of nodes and data sources were experimented but only the most significant results for a higher traffic load of 50 sources and a reduced network size of 50 nodes are reported. An attempt to simulate above 100 nodes was aborted due to slow speed and large memory requirement of the simulator. A traffic load of 50 sources is also the highest that could be injected without severely congesting the network for an otherwise meaningful comparison.

Table 4 summarizes the previous results for a 100 node model with 40 sources, under highest mobility (pause time = 0s). The overhead results are normalized to the total number of data packets delivered, i.e. the number of overhead packets transmitted per data packet successfully delivered to the destination. This gives a measure of the protocol efficiency that can be directly compared between different experiments in this section. The percentage improvement of DSR+S over DSR for each metric is also shown.

Effect of increased traffic load is illustrated in Table 5a. Both protocols incurred a higher routing overhead as more routes are discovered and maintained for the greater number of data sources. Route discovery, end-to-end delay and packet delivery performance deteriorated due to congestion and collisions caused by the higher data and control traffic. However, DSR+S still performed significantly better across all metrics. The margin of improvement is higher with larger number of sources.

With a smaller network size of 50 nodes, the overhead generated is less and so is the margin of performance improvement (Table 5b). Packet delivery performance seems relatively unchanged, which suggests that the network is still sufficiently connected even though the number of nodes is reduced. An interesting observation though, is the increase in end-to-end delay, which contradicted our earlier intuition that less overhead and hence lower congestion, reduce the end-to-end delay. We checked that both 50 and 100 node models have similar average path length of approximately

Table 4: Summary of results for a 100 node model with 40 sources

| Performance metrics | DSR | DSR+S | % Improvement |
|---|---|---|---|
| Route discovery overhead (normalized) | 2.92 | 1.83 | 37.3 |
| Total routing overhead (normalized) | 4.19 | 2.62 | 37.5 |
| Route discovery latency (ms) | 236 | 41.7 | 82.3 |
| End-to-end delay (ms) | 334 | 104 | 68.9 |
| Packet delivery ratio (%) | 85.7 | 90.9 | 6.07 |

Table 5: Effect of changing network size and traffic load

| Performance metrics | DSR | DSR+S | % Improvement |
|---|---|---|---|
| Route discovery overhead (normalized) | 7.42 | 2.45 | 67.0 |
| Total routing overhead (normalized) | 11.2 | 4.31 | 61.5 |
| Route discovery latency (ms) | 1387 | 589 | 57.5 |
| End-to-end delay (ms) | 1641 | 649 | 60.5 |
| Packet delivery ratio (%) | 52.6 | 76.1 | 44.7 |

(a) Traffic load: 50 sources. Network size maintained at 100 nodes

| Performance metrics | DSR | DSR+S | % Improvement |
|---|---|---|---|
| Route discovery overhead (normalized) | 1.12 | 1.0 | 10.7 |
| Total routing overhead (normalized) | 1.63 | 1.5 | 7.98 |
| Route discovery latency (ms) | 55.6 | 37.6 | 32.4 |
| End-to-end delay (ms) | 567 | 539 | 4.94 |
| Packet delivery ratio (%) | 87.7 | 90.1 | 2.74 |

(b) Network size: 50 nodes. Traffic load maintained at 40 sources

4 hops. Thus, it is not due to an increase in path length that increases the end-to-end delay.

Instead, buffering delay at the source nodes for route discovery is the main factor. This is based on our finding that cache hit rate of the 50 node model is much lower than that of 100 node model by an order of magnitude (Table 6). A cache hit refers to a cache access that successfully finds the requested route. This higher cache misses means that more data packets must wait at the source (in a buffer) for a route to their destination and not send immediately, leading to a higher delivery delay. The lower cache hit rate of the 50 node model is attributed to the reduced route diversity, or fewer number of routes that can exist with a reduced number of nodes.

Table 6: Cache hit rates with 40 sources and zero pause time

| Number of nodes | Cache hit rates | |
|---|---|---|
|  | DSR | DSR+S |
| 50 | 0.0286 | 0.0294 |
| 100 | 0.3431 | 0.2472 |

## VI. Conclusions

In this paper, we reported a phenomenon that occurs during a route discovery flood, for causing an intermediate node to notice a RREP for a RREQ that it receives only later, due to various effects on the propagation of RREQ, such as channel contention, path length, packet collision and mobility. The observation of the phenomenon inspires an approach for suppressing non-optimal RREP from intermediate nodes, thus reducing the number of Cached RREP, which is known to be a major source of routing load for DSR.

The proposed approach is generally simple in concept and implementation, and is found to be effective in improving not only the efficiency of DSR in terms of a lower routing load, but also its data delivery performance in terms of a higher delivery success rate and lower delay, particularly in more stressful environments such as under high mobility or traffic load.

It may be interesting to see how the proposed scheme may also enhance other protocols such as [14, 15] (AODV/DYMO with path accumulation)

which are known to share similar characteristics as DSR. We will leave this as future work.